\documentstyle{l-aa}     % LaTeX A&A  Standard Fonts
\newcommand{\mo}    {\hbox{${\rm M}_{\odot}$}}

\newcommand{\kms}   {\hbox{${\rm km\,s}^{-1}$}}

\newcommand{\cmsq}  {\hbox{${\rm cm}^{-2}$}}
\newcommand{\cmcb}  {\hbox{${\rm cm}^{-3}$}}

\newcommand{\jone}  {$J$=1$-$0}                         % J=1-0
\newcommand{\jtwo}  {$J$=2$-$1}                         % J=2-1
\newcommand{\jowt}  {$J$=2$\leftarrow$1}                % J=1-2
\newcommand{\ergs}   {\hbox{${\rm erg\,s}^{-1}$}}
\newcommand{\K}      {\hbox{\rm\thinspace K}}
\newcommand{\pcs}      {\hbox{\rm\thinspace pc}}
\begin{document}
\topmargin=2.0cm
\thesaurus{ 09.13.2;  % Interstellar medium : molecules
            11.02.2;  % Galaxies            : BL Lac objects individual
	    11.09.4;  % Galaxies            : ISM
            11.17.1   % (Galaxies)          : quasars         : absorption lines
            13.18.3;  % Sources as func...  : Radio continuum : galaxies
	         }
%%%%%%%%%%%%%%%%%%%%%%%%%%%%%%%%%%%%%%%%%%%%%%%%%%%%%%%%
%
 \title{A search for molecular absorption in the tori of active galactic nuclei}
%
%%%%%%%%%%%%%%%%%%%%%%%%%%%%%%%%%%%%%%%%%%%%%%%%%%%%%%%%
%
   \author{M.J.~Drinkwater\inst{1}, F.~Combes\inst{2},  T.~Wiklind\inst{3}}
   \offprints{M.~Drinkwater, mjd@aaocbn.aao.gov.au}
   \institute{Anglo-Australian Observatory, Coonabarabran, NSW 2357, Australia
   \and
              DEMIRM, Observatoire de Paris, 61 Av. de l'Observatoire,
              F--75014 Paris, France
   \and
              Onsala Space Observatory, S--43992 Onsala, Sweden
}
   \date{Received date; Accepted date}
   \maketitle
   \markboth
{Drinkwater et al.: Search for molecular absorption in the tori of AGN}
{Drinkwater et al.: Search for molecular absorption in the tori of AGN}
%%%%%%%%%%%%%%%%%%%%%%%%%%%%%%%%%%%%%%%%%%%%%%%%%%%%%%%%
%
\begin{abstract}

We describe a search for molecular absorption at millimetre
wavelengths associated with dusty
molecular tori in active galactic nuclei (AGN).
The sample observed consists of 11
flat-spectrum radio sources known to have red optical to infra-red
colours plus two steep-spectrum narrow-line radio galaxies. Spectra of
the sources were obtained in the 3-, 2- and 1.3-millimetre bands at
frequencies corresponding to common molecular transitions of CO,
HCO$^+$, HCN and CS at the AGN redshift. The observations were thus
sensitive to absorption taking place either in dusty molecular tori
surrounding the AGN nucleus, or in molecular clouds in the AGN host
galaxy.

No absorptions were detected in any of the sources. We calculated
upper limits to the column density in molecular absorption, using an
excitation temperature of 10\K, to be
$N_{CO} < 10^{15} - 10^{16} \cmsq$, equivalent
to hydrogen columns of order $N_H < 10^{19} - 10^{20} \cmsq.$ These
limits are significantly lower than the values ($N_H \approx (2 -
6)\times 10^{21} \cmsq$) that might be expected if the red colours of
these sources were due to dust absorption at the quasar redshift
as suggested by Webster et al.\ (1995).
Should the excitation temperature of the molecular transitions be
higher than 100K, the upper limits to the H$_2$ column densities
would be greater than those derived from the red colours.

To explain the lack of molecular absorption we conclude that either the
optical extinction takes place outside the host galaxy (along the line
of sight), or the excitation temperature of the molecular transitions
is very high, or the obscuration is not associated with significant
amounts of cold molecular gas. It is quite possible that the hard X-ray
flux from the central source of these AGN is strong enough to
photo-dissociate the molecules.

\end{abstract}
%________________________________________________________________
\section{Introduction}

In the unified models of active galactic nuclei (AGN), the central
``engine'' is surrounded by a parsec-scale geometrically and optically
thick dust torus. The Broad Line Region (BLR) and a non--thermal
continuum source are situated inside the torus, whereas the Narrow Line
Region (NLR) is outside. Many observed characteristics and apparent
differences of AGN can then be ascribed to orientation effects (see
Antonucci 1993).

Although the unified model is hotly debated, it is clear that at least
some AGN are surrounded by tori. That opaque tori can block the BLR
from direct view has been demonstrated by spectropolarimetry, where the
polarized flux spectra show broad permitted lines in AGN which normally
are characterized by forbidden narrow lines only, i.e. a hidden BLR
(cf. Antonucci \& Miller 1985).

In nearby AGN there is evidence for the presence of molecular gas that
may be associated with dusty tori. Very Long Baseline Array
observations of H$_{2}$O maserspots in the nucleus of NGC\,4258
(Miyoshi et al.\ 1995) strongly suggest a circumnuclear molecular torus
with Keplerian rotation. However the extent of the maser emission
region is only 0.13\pcs\ and the disk is not geometrically thick.  On
much larger scales than proposed for tori, Planesas et al.\ (1991)
found a circumnuclear molecular ring with a radius of 130\pcs\ in
NGC\,1068, a prototypical Seyfert 2 galaxy, and in NGC\,4945, Bergman
et al.\ (1992) inferred a thick torus consisting of a large number of
small but dense molecular clouds.  A circumnuclear ring has also been
seen in Centaurus A, with a radius of $\sim100$\pcs\ (Israel et
al.\ 1990, Rydbeck et al.\ 1993).

These observational data clearly show the existence of molecular gas
close to the central engine of AGN, but the dimensions are rather
extreme and the masses are quite small. The rather large torus in
Centaurus A only contains $\sim10^{6}$\,\mo\ of molecular gas (Rydbeck
et al.\ 1993). The physical conditions in the molecular tori, such as
density and temperature, are largely unknown. The small molecular mass
combined with the low beam filling factor\footnote{In the most nearby
AGN, the tori extend $\sim3-5$ arcseconds, whereas a millimeter
telescope beam is of order 10 arcseconds or more.} of nuclear tori
means that observations of CO {\em emission} from more distant sources
will be beyond the capacity of present and future millimetre-wave
telescopes.

Molecular {\em absorption}, however, is much easier to detect and can
be observed at any distance as long as the background source remains
unresolved and has a reasonably strong continuum flux at millimeter
wavelengths. The utility of mm--wave molecular absorption lines in
deriving detailed properties of molecular gas at large distances has
been demonstrated by the observations of 38 different molecular
transitions in four absorption systems at redshifts $z=0.25-0.89$
(Wiklind \& Combes 1994, 1995, 1996a, 1996b).  The high sensitivity of
molecular absorption line observations means that molecular species
much rarer than CO can be observed. For instance, HCO$^+$, HCN, HNC, CS
and CN have been observed, as well as the isotopic variants $^{13}$CO
and C$^{18}$O (Combes \& Wiklind 1995) as well as H$^{13}$CO$^+$
(Wiklind \& Combes 1996a). Hence we have the tools to
derive detailed physical and chemical properties of the molecular
component of the interstellar medium at high redshifts, something
which is impossible by other means.  Molecular absorption lines can
therefore be used to search for molecular tori very close to the
presumed black holes in distant AGN.

The main problem in observing molecular gas close to the centre of AGN
is to select the best candidates. Many quasi-stellar object (QSO)
surveys have selection criteria which favour objects which do not
suffer substantial extinction (blue colours, UV--excess, etc). A better
approach is to use surveys compiled at wavelengths which are unbiased
concerning extinction. Radio surveys are suitable; indeed several
radio--loud QSOs (i.e.\ quasars) are known to be optically faint, but
are bright in the near--infrared. These sources are most likely
``optically quiet'' because of extinction.  Webster et al.\ (1995) have
recently discovered a population of radio--loud quasars that are 2--4
magnitudes redder in optical to infra-red $B_J-Kn$
colours\footnote{$B_J$ is a blue photographic band formed by IIIa-J
emulsion with a GG385 filter and $Kn$ is an infra-red band covering
2.0-2.3$\mu$m} than samples of optically selected QSO samples. The
distribution of emission line equivalent widths indicates that the red
colours are not due to an intrinsically red spectrum from the central
continuum source, so they conclude that the red colours are caused by
dust absorption. There are three possible sources of the extinction:
\begin{enumerate}

\item An intervening galaxy (which seems to be the case
for B0218+357 and PKS~1830$-$211).

\item Molecular clouds in the AGN host galaxy (PKS~1413+135 and B31504+377).

\item A molecular torus obscuring the line of sight to the nucleus.

\end{enumerate}

In this paper we present data from a search of molecular absorption
directly associated with the AGN themselves, i.e. cases 2 and 3,
concentrating on sources from Webster et al.\ (1995) that are known to
be optically red.  We did not detect any absorption. This does not
necessarily imply that the first case is correct, but it does put very
stringent limits to the amount of molecular material in the AGN host
galaxy or torus along the line of sight.  In Section~2 we describe the
sources observed and in Section~3 we present the results of our
observations: high signal-to-noise spectra which did not reveal
absorption in any of the sources.  In Section~4 we derive upper limits
to the molecular absorption columns in our sources and compare these to
estimates of the absorption required by the optical reddening.  A
discussion of possible scenarios for explaining the lack of molecular
gas along the line of sight to heavily obscured AGN and the
consequences for the unified model are discussed in Section~5.

\section{The Sample}

As our intention was to observe objects known to be obscured, we
selected 11 red ($B_J-Kn\geq 4$) sources from the Webster et
al.\ (1995) sample. The sample is fully listed and defined in
Drinkwater et al.\ (1996); basically it consists of flat-spectrum radio
sources brighter than 0.5 Jy at 2.7 GHz.  In the unified model,
flat-spectrum sources are most likely seen with the torus ``face-on''
giving an unobstructed view of the central source of the AGN.  This
would tend to select against cases where we might expect to have
significant absorption by the torus. This is to some degree borne out
by the fact that broad permitted emission lines are seen in many of the
optical spectra (see Table~1). There is probably some range of
orientations in the sample however, e.g.\ the radio--galaxy
PKS~0521$-$365 which is an intermediate case with a spectral index of
$-0.49$ (at the steep limit of our sample), and a radio morphology of
two lobes plus a compact core.

We observed two additional sources which are much more likely to have
edge--on configuration; the powerful narrow--line radio galaxies
Cygnus~A and Hydra~A. We do not have such complete information about
these sources, but Cygnus~A is red in $B_J-Kn$ and Hydra~A is
steep-spectrum by our definition, and both have extended radio
structure indicative of edge-on configurations (e.g.\ Carilli et
al.\ 1994 for Cygnus A and Morganti et al.\ 1993 for Hydra A).

The sources we observed, which are listed in Table~1, display a range
of properties relating to their orientation, ranging from the compact
core-dominated flat-spectrum quasars to more extended steep--spectrum
radio--galaxies.

%--------------------------Table 1--------------------------------------------
\begin{table*}
\begin{flushleft}
\caption[]{Parameters of the Sources Observed}
\scriptsize
\begin{tabular}{lrrrrllllll}
\hline
 & & & & & & & \\
  Source & RA(B1950) &Dec(B1950)& $\alpha$& ID& $z_{\rm e}$&$\Delta v^{a)}$& 
$B_J$&$B_J-Kn^{b)}$& $N_H$Gal$^{c)}$&$N_H$X-ray \\
       &  hours& degrees&      &   &  & \kms  & mag& mag& $10^{20}$\cmsq& 
$10^{20}$\cmsq \\
 & & & & & & & \\
\hline
 & & & & & & & \\
  PKS~0113$-$118&   1:13:43.22&  $-$11:52:04.5&    0.09& Q & 0.672 &   5200& 
19.42& 4.1&  3.1& -   \\
  PKS~0422+004  &   4:22:12.52&      0:29:16.7&    0.35&BL & 0.310 &      -& 
16.19& 3.7&  8.7& -   \\
  PKS~0438$-$436&   4:38:43.18&  $-$43:38:53.1&    0.12& Q & 2.852 &   1700& 
19.08& 2.7&  1.8& 6.5,86$^{e)}$\\
  PKS~0521$-$365&   5:21:12.99&  $-$36:30:16.0& $-$0.49& G & 0.0552&   1800& 
16.74& -  &  2.7& 4.5$^{d)}$\\
  PKS~0537$-$441&   5:37:21.00&  $-$44:06:46.8& $-$0.02& Q & 0.894 &   2400& 
15.45& 3.9&  3.1& (3.1)$^{d)}$\\
  Hydra A       &   9:15:41.45&  $-$11:53:08.3& $-$0.95& G & 0.0538&$<$1500& 
13.7 & -  &  4.6& -   \\
  PKS~1213$-$172&  12:13:11.67&  $-$17:15:05.3& $-$0.06& X & -     &      -& 
-    & -  &  4.4& -   \\
  PKS~1548+056  &  15:48:06.93&      5:36:11.3&    0.28& Q & 1.422 &   4400& 
18.45& 4.3&  4.7& -   \\
  PKS~1555+001  &  15:55:17.69&      0:06:43.5&    0.21& Q & 1.77  &   4700& 
22.12& 6.6&  6.9& -   \\
  PKS~1725+044  &  17:25:56.34&      4:29:27.9&    0.71& Q & 0.296 &   4500& 
18.20& 4.4&  7.3& (21)$^{d)}$\\
  Cygnus A      &  19:57:44.45&     40:35:46.1&    -   & G & 0.0561&$<$1500& 
17.0 & 4.4&  41.& 820 $^{f)}$ \\
  PKS~2223$-$052&  22:23:11.08&   $-$5:12:17.8& $-$0.14& Q & 1.404 &      -& 
17.12& 3.7&  5.1& (6.2)$^{d)}$\\
  PKS~2329$-$162&  23:29:02.40&  $-$16:13:30.8&    0.08& Q & 1.155 &   2900& 
20.87& 4.3&  3.6& -   \\
  & & & & & & \\
\hline
\end{tabular}
\ \\
Note: The following data are from Drinkwater et al.\ (1996) (except Cygnus~A 
and Hydra~A from NED): RA and Dec are the accurate radio source positions;
$\alpha$ is the spectral index ($S_\nu\propto\nu^\alpha$) measured from 2.7 
to 5.0 GHz; ID identifies the source as Quasar, BLlac, Galaxy or X=near a 
bright star; $z_{\rm e}$ is the redshift.\\
a)\ Line widths (FWHM) of permitted lines estimated from the spectra cited in 
Drinkwater et al.\ (1996). \\
b)\ $Kn$ magnitudes from Webster (private communication) except for 
0438$-$436, Elston (private communication). \\
c)\ Galactic $N_H$ column density estimated from Stark et al.\ (1992). \\
d)\ X-ray $N_H$ column density estimated from ROSAT hardness ratios; sources 
with no significant excess in parentheses (Brinkmann, private communication). \\
e)\ X-ray $N_H$ column density from Elvis et al.\ (1994a), two values at 
rest and quasar redshifts respectively. \\
f)\ X-ray $N_H$ column density from Arnaud et al.\ (1987).
\end{flushleft}
\end{table*}
% PKS~0438-436 Ks=16.38 mag from Richard Elston on CTIO 1m 1995 Nov 18/19
%----------------------End Table 1--------------------------------------------

\section{Observations}

The observations were made with the IRAM 30--m telescope at Pico
Veleta in Spain, during the period 1995 June 10--16 and on the SEST
15--m telescope at La Silla in Chile on a number of runs in 1993
January and 1995 June--July. We used the 3--, 2-- and 1.3--mm SIS
receivers, tuned to the redshifted frequencies of the molecular
transitions in question (see Table\,2).  The data for most of the
sources were presented previously by Wiklind \& Combes (1996b) where
full details of the observing setup are given, although we note in
particular that the bandpass of the spectra was 500 MHz for the IRAM
data and 1000 MHz for the SEST data. In this paper we present new
measurements of PKS~0438$-$436, PKS~2223$-$052, Cyg\,A and Hydra\,A.

% For 1555+001 with the poor redshift Baldwin et al quote:
% 3385 1.784 Ly-a   \_ mean z=1.77
% 4280 1.761 CIV    /

No absorption was detected in any of the spectra. The sources were
bright at millimetre-wavelengths so high signal-to-noise spectra were
obtained (see the spectra of Cygnus~A in Figure~1) which allowed us to
place upper limits on the molecular absorption columns. Details of the
spectra and the derived upper limits are given in Table\,2. Before
discussing the absorption limits in the next section it is necessary to
confirm that the observations sampled a large enough range in redshift
to include the emission redshift of each source. Note that this is
not guaranteed for two of the sources: PKS~0422+004 has no published
reference and for PKS~1555+001 the measurement is not precise (Baldwin
et al.\ 1981).  In Table~2 we list the best available optical redshift
for each source along with an error if available. The cases where no
error is given are all of reasonable quality so we can assume an error
of $\Delta z = 0.002$.  Then for each of the transitions involved we
have calculated the range of redshifts  ($z_{\rm observed}$) covered on
the basis of the central frequency observed and the appropriate
bandpass.  Note that in some cases a second overlapping frequency range
was also observed.  We find that in all cases except the two mentioned
above, the spectral range is sufficient to have included the emission
redshift.

\begin{figure}
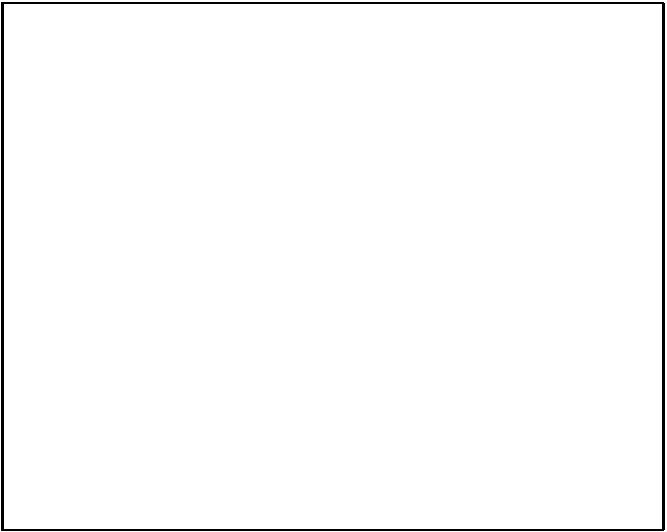

\picplace{7.0cm}
%\special{psfile=mmfig.ps hoffset=-50 voffset=213 vscale=40 hscale=40 angle=270}
\caption[]{Low resolution ($\Delta v=10.8$\,\kms) spectra of 
CO(\jone), HCO$^+$(\jowt) and CO(\jtwo) towards Cygnus~A.
The velocity scale is heliocentric and centered at $z_{\rm e}=0.0566$.}
\end{figure}

\begin{table*}
\begin{flushleft}
\caption[]{Molecular tori not detected in absorption}
\scriptsize
\begin{tabular}{llrrlrrrlll}
\hline
 & & & & & & & & & & \\
{Source}                       &
{$z_{\rm e}$}                  &
{Transition}                   &
{$\nu_{\rm obs}$}              &
{$z_{\rm observed}$}           &
{T$_{\rm cont}$}               &
{t$^{a)}$}                     &
{$\delta$v}                    &
{$\sigma_{\rm rms}$}           &
{$\int \tau_{\nu}\,dv^{b)}$}   &
{$N_{\rm tot}$}                \\
{ } & { } &  { }         &
{GHz}                          &
                               &
{K}                            &
{  }                           &
{${\rm km\,s}^{-1}$}           &
{mK}                           &
{${\rm km\,s}^{-1}$}           &
{${\rm cm}^{-2}$}              \\ 
 & & & & & & & & & & \\
\hline
 & & & & & & & & & & \\
%----------------------------------------------------------------------------
PKS~0113$-$118  & 0.672 &     HCO+(2--1)   & 
106.68  &$0.672^{+0.004}_{-0.010}$&  0.090 & I & 2.8 & 5.6 & 0.37 & $< 2 
\times10^{12}$      \\   
WC96$^{d)}$  &   	& 	 CO(2--1)   & 
137.88  &$0.672{\scriptstyle\pm0.003}$&  0.090 & I & 2.2 & 8.9 & 0.48 & $< 1 
\times10^{15}$      \\   
	& 	& 	 CO(3--2)   & 
206.81  &$0.672{\scriptstyle\pm0.002}$&  0.040 & I & 1.8 & 13. & 1.89 & $< 1
 \times10^{16}$      \\   
&&&&&&&&&&\\
%----------------------------------------------------------------------------
PKS~0422+004  & 0.310 ? &     CO(1--0)   & 
87.993 &$0.310{\scriptstyle\pm0.008}$&  0.035 & S & 12. & 5.2 & 4.23 & $<2 
\times10^{16}$   \\ 
WC96	& 	& 	 HCN(2--1)  & 
135.31 &$0.310{\scriptstyle\pm0.005}$&  0.030 & S & 9.6 & 6.9 & 5.92 & $<3 
\times10^{13}$   \\   
	& 	& 	 HCO+(2--1)  & 
136.16 &$0.310{\scriptstyle\pm0.005}$&  0.030 & S & 9.6 & 6.9 & 5.92 & $<3 
\times10^{13}$   \\   
	& 	& 	 HNC(2--1)  & 
138.41 &$0.310{\scriptstyle\pm0.005}$&  0.030 & S & 9.6 & 6.9 & 5.92 & $<4 
\times10^{13}$   \\   
&&&&&&&&&&\\
%----------------------------------------------------------------------------
PKS~0438$-$436  & $2.852\pm0.003$      & CO(3--2)   & 
89.770  &$2.852{\scriptstyle\pm0.021}$&  0.045 & S & 2.3 & 3.9 & 0.44 & $< 2 
\times10^{15}$  \\
&&&&&&&&&&\\
%----------------------------------------------------------------------------
PKS~0521$-$365  & 0.055     & HCO+(1--0)   & 
84.538  &$0.0552{\scriptstyle\pm0.006}$&  0.200 & S & 9.6 & 5.8 & 0.57 & $< 6
 \times10^{12}$   \\
WC96    &                              &        &   &     &     &      & \\
&&&&&&&&&&\\
%----------------------------------------------------------------------------
PKS~0537$-$441  & 0.893     & HCN(2--1)  & 
93.591 &$0.894{\scriptstyle\pm0.010}$&  0.250 & S & 12. & 10. & 1.00 & $<6 
\times10^{12}$  \\
WC96	& 	& 	 HCO+(2--1) & 
94.179 &$0.894{\scriptstyle\pm0.010}$&  0.250 & S & 12. & 10. & 1.00 & $<5 
\times10^{12}$  \\
	& 	& 	 HNC(2--1) & 
95.736 &$0.894{\scriptstyle\pm0.010}$&  0.250 & S & 12. & 10. & 1.00 & $<7 
\times10^{12}$  \\
	& 	& 	 CS(4--3) & 
103.46 &$0.894{\scriptstyle\pm0.009}$&  0.250 & S & 12. & 10. & 1.00 & $<3 
\times10^{13}$  \\
	& 	& 	 CS(5--4) & 
129.32  &$0.894{\scriptstyle\pm0.007}$&  0.170 & S & 15. & 6.5 & 1.19 & $< 6 
\times10^{13}$ \\  
&&&&&&&&&&\\
%----------------------------------------------------------------------------
Hydra~A  & $0.0538\pm0.0001$ &     CO(1--0)   & 
109.39  &$0.0538{\scriptstyle\pm0.002}$&  0.060 & I & 2.7 & 9.4 & 1.01 & $< 6 
\times10^{15}$   \\   
	& 	& 	 HCO+(2--1)   & 
169.27  &$0.0538{\scriptstyle\pm0.002}$&  0.050 & I & 7.2 & 13. & 5.28 & $< 2 
\times10^{13}$ \\ 
	& 	& 	 CO(2--1)   & 
218.77  &$0.0538{\scriptstyle\pm0.001}$&  0.040 & I & 6.8 & 10. & 4.71 & $< 1 
\times10^{16}$    \\ 
&&&&&&&&&&\\
%----------------------------------------------------------------------------
PKS~1213$-$172  &$^{c)}$ &  3mm   & 
80--104& &  0.080 & S & 15. & 7.2 & 0.20 & \\  % 0. & $< $ \\
WC96	& 	& 	 2mm  & 
128--153& &  0.070 & S & 13. & 6.4 & 0.20 & \\  % 0. & $< $ \\  
&&&&&&&&&&\\
%----------------------------------------------------------------------------
PKS~1548+056  & $1.422\pm0.001$      & CO(2--1)   &  
95.185  &$1.422^{+0.006}_{-0.016}$&  0.250 & I & 3.1 & 10. & 0.25 & $< 8 
\times10^{14}$  \\
WC96	& 	& 	 CO(3--2)   & 
142.77  &$1.422{\scriptstyle\pm0.004}$&  0.160 & I & 2.1 & 10. & 0.28 & $< 1 
\times10^{15}$     \\   
&&&&&&&&&&\\
%----------------------------------------------------------------------------
PKS~1555+001  & $1.77\pm0.01$      & HCO+(3--2)   & 
96.591  &$1.770{\scriptstyle\pm0.007}$&  0.110 & I & 3.1 & 9.1 & 0.56 & $< 3 
\times10^{12}$   \\
WC96	& 	& 	 CO(4--3)   & 
166.44  &$1.770{\scriptstyle\pm0.004}$&  0.100 & I & 1.8 & 25. & 1.25 & $< 2 
\times10^{16}$     \\   
&&&&&&&&&&\\
%----------------------------------------------------------------------------
PKS~1725+044  & 0.296      & CO(1--0)   & 
88.943  &$0.296{\scriptstyle\pm0.004}$&  0.120 & I & 3.4 & 11. & 0.69 & $< 4 
\times10^{15}$   \\
WC96	& 	& 	 HCO+(2--1)   & 
137.63  &$0.296{\scriptstyle\pm0.002}$&  0.100 & I & 2.2 & 12. & 0.60 & $< 3 
\times10^{12}$     \\   
&&&&&&&&&&\\
%----------------------------------------------------------------------------
Cygnus~A  & $0.0561\pm0.0001$ &      CO(1--0)   & 
109.10  &$0.0566^{+0.002}_{-0.007}$&  0.120 & I & 2.7 & 4.9 & 0.23 & $< 1 
\times10^{15}$  \\
	& 	& 	 HCO+(2--1)   & 
168.82  &$0.0566{\scriptstyle\pm0.002}$&  0.080 & I & 1.8 & 9.3 & 0.48 & $< 2 
\times10^{12}$ \\ 
	& 	& 	 CO(2--1)   & 
218.19  &$0.0566{\scriptstyle\pm0.001}$&  0.070 & I & 1.7 & 6.5 & 0.35 & $< 1 
\times10^{15}$     \\   
&&&&&&&&&&\\
%----------------------------------------------------------------------------
PKS~2223$-$052  & $1.404\pm0.003$      & CO(2--1)   & 
95.897  &$1.404{\scriptstyle\pm0.004}$&  0.285 & I & 3.1 & 4.2 & 0.09 & $< 3 
\times10^{14}$  \\
WC95$^{e)}$&                          &        &   &     &     &      & \\
&&&&&&&&&&\\
%----------------------------------------------------------------------------
PKS~2329$-$162  & 1.155      & CO(2--1)   & 
106.98  &$1.155^{+0.015}_{-0.005}$&  0.080 & I & 2.8 & 6.6 & 0.50 & $< 2 
\times10^{15}$   \\
	& 	& 	 CO(3--2)   & 
160.46  &$1.155{\scriptstyle\pm0.003}$&  0.060 & I & 1.9 & 12. & 0.97 & $< 5 
\times10^{15}$    \\   
&&&&&&&&&&\\
\hline
\end{tabular}
\ \\
a) Telescope: I=IRAM; S=SEST. \\
b)  $\tau_{\nu} = -\ln (1 - 2\sigma_{\rm rms}/T_{\rm cont})$.\\
c) This source is next to a very bright star for which reason the emission
redshift is unknown; a particular effort was made to search a large part of
2- and 3-mm bands for any absorption features; none were detected. \\
d) WC96: Wiklind \& Combes 1996b \\
e) WC95: Wiklind \& Combes 1995
%----------------------End Table 3bis----------------------------------------
\end{flushleft}
\end{table*}

\section{Analysis of Absorption Limits}

\subsection{Molecular Absorption Upper Limits}

We used the non--detection of absorption in our data to place upper
limits on the column density in the various molecules. These are shown
in Table~2. The calculations were made in the same way as in Wiklind \&
Combes (1996b), except that we have taken a slightly less conservative
limit of 2--sigma, based on the noise in a single spectral element of
width $\delta v$ (indicated in column 8 of Table~2) and assuming an
excitation temperature of 10\K.

The upper limits on absorption at $z = z_{em}$ in the various sources
were of the order $N_{CO} < 10^{15}-10^{16}\cmsq$ (similar results were
obtained for the other molecular species using the relative abundances
e.g.\ $N_{CO}/N_{HCN}\approx N_{CO}/N_{HCO}\approx10^4$).  These column
densities were derived assuming an excitation temperature of 10\K. This
is a value typical for molecular gas in a variety of different
environments, both in our own and nearby galaxies. The excitation may
be quite different in tori close to an AGN, and the implication of this
will be discussed in Section~5.  If we assume that the local ratio of CO
to H$_2$ of $\approx 10^{-4}$ (Dickman, 1978) can be applied, these
limits correspond to a hydrogen column density of
$$N_H   <  10^{19} - 10^{20} \cmsq.$$

\subsection{Expected Absorption from Optical Data}

We now consider what hydrogen column we might expect from the optical
reddening proposed by Webster et al.\ (1995). They find a mean colour
of $B_J-Kn\approx 2.5$ mag for the optical Large Bright QSO Survey
(LBQS) compared to the redder colours of flat--spectrum radio quasars.
For the sample observed here the colour range is $B_J-Kn = 3.7 - 6.6$
mag. If we assume all the difference is due to absorption in $B_J$,
then $A_B \approx 1 - 4$ mag. If the absorption follows a $1/\lambda$
dust law then $A_B / A_V \approx 1.33$, so $A_V = 1 - 3$ mag. We
can use the Bohlin et al.\ (1978) results for the Solar neighbourhood
to convert this absorption to a hydrogen column density ($E(B-V)\approx
A_V/3$ and $N(HI + H2) \approx E(B-V) \times 5.8\times 10^{21} \cmsq$).
This gives
$$   N_H \approx (2 - 6)\times 10^{21} \cmsq.$$
 
If all the reddening is dust {\em at the quasar redshift} it implies a
hydrogen column density 2 orders of magnitude larger than the upper
limit implied by the CO data\footnote{These values both rely on the use
of relations derived for the Solar neighbourhood. These may not be
valid for the galaxies studied, but the errors are unlikely to change
the ratio of the derived column densities significantly, since both
the dust-to-gas ratio and the heavy-element molecules are proportional
to the metallicity}.  This could
imply that the absorption is not taking place in the AGN host galaxy or
that the excitation temperature is at least an order of magnitude
higher than what is normally found in molecular gas; alternatively the
dust may not have any associated molecular absorption: see the
discussion in Section~5.

\subsection{X-ray Absorption}

An independent measure of the $N_H$ absorption column for some of our
sources is given by detections of X-ray absorption. Elvis et
al.\ (1994a) report a significant X-ray absorption for the source PKS
0438$-$436, well in excess of foreground Galactic $N_H$. They measure
excess X-ray absorption equivalent to a column of $N_H=6.5\times
10^{20} \cmsq$, but the column needed to produce the required absorption
increases with the absorption redshift because the photons gain energy
according to $E\propto(1+z)$ (see Elvis et al.\ 1994a). In the case of
this source, the absorption becomes much larger at $z=2.8$ giving
$$N_H=8.6\times 10^{21} \cmsq.$$

We searched the ROSAT public sky survey catalogue for detections
of other sources in our sample; 4 were found (including PKS~0438$-$436) that
had good enough data to estimate the total absorption column
using the hardness ratio method (see Schartel et al.\ 1992). These
estimates were made by Brinkmann (private communication) and are shown
in Table~1. In two of the four cases (PKS~0438$-$436 and 0521$-$365) the values
were significantly in excess of that expected from the Galaxy. We also 
note that a very large amount of absorption was reported for Cygnus~A
by Arnaud et al.\ (1987), although that is a rather special case because
of the complex cluster environment of the source as well as the large
Galactic column density.

Although we only have X-ray data for a very small number of our
sample, the inferred absorption columns are consistent with the large
amounts of absorption we obtain for the optical reddening
hypothesis. (Note that a similar absorption column of $N_H=9\times
10^{21} \cmsq$ was obtained for the steep-spectrum ``red quasar''
3C~212 by Elvis et al.\ 1994b.)  These values are in excess of the
upper limits we infer above from the millimetre 
data of $N_H<10^{19} \cmsq$, but are not inconsistent with
the lack of absorption at the quasar redshift because the redshift of
the X-ray absorption is not known.
 
\section{Discussion}

A previous attempt to detect CO absorption towards Cygnus A failed
(Barvainis \& Antonucci 1994) and our results put even stronger limits
on this non-detection. If we consider $3\sigma$ upper limits within
channels of 6\kms, we find that the optical depth in CO $\tau_{CO}$ is
lower than 0.08, while Barvainis \& Antonucci found a limit of
$\tau_{CO}<0.6$.  If broader lines are considered, the limits are even
stronger, evolving as the square root of the line-width.  This result
is somewhat disappointing since the nucleus of Cygnus A exhibits strong
extinction at optical wavelengths as well as a significant column
density of hydrogen as show by the X-ray results.  Indeed, there is an
appreciable HI column density, as observed through HI absorption
(Conway \& Blanco 1995): $2.54\times10^{22}\cmsq$, if T$_{spin}
\approx$ 1000K. The absorption appears to come from nearby the nucleus,
because of its broad width (280\kms),
but not closer than $\sim$40\pcs\ as deduced from the lack of free--free
absorption at 1.4\,GHz (Conway \& Blanco 1995; Maloney 1996).
None of our other sources
revealed any molecular absorption either, even PKS~0438$-$436 which
also has significant X-ray absorption.

There are several possible causes for the non--detections. 
The most reasonable are:
\begin{enumerate}

\item 
The material causing the optical and X-ray absorption may not have any
molecular gas associated with it because of photo-dissociation by the
ionizing flux from the AGN nucleus. This is supported by the fact that
appreciable HI column density is detected in absorption in Cygnus A.
This effect was considered by Maloney et al.\ (1994), who showed that
for CO to be present in significant amounts the ``effective'' X-ray
ionization parameter
$$\xi_{eff}=1.1\times10^{-2}L_{44}/(n_9r^2_{pc}N^{0.9}_{22})$$
must satisfy the condition $\xi_{eff} \le 5\times10^{-3}$, where
$L_{44}$ is the hard (2--10\,keV) X-ray luminosity in units of
$10^{44}\ergs$, $n_9$ is the density in units of $10^9\cmcb$, $r_{pc}$
is the distance from the nucleus in pc, and $N_{22}$ is the hydrogen
column density from the source. 
In the case of Cygnus A with an X-ray
luminosity of $10^{44}\ergs$ and radii of order 1\pcs, Maloney et
al.\ find that the condition can be satisfied if the density is as high
as order $10^9\cmcb$.  This is ``not implausible'' so
photo-dissociation might not be important at small radii. 
If so, it would not be important at larger radii as well, since 
the condition for non-photodissociation is satisfied for radii larger
than a critical one, given the column density imposed by the extinction
and HI absorption. In the case of an isobaric model, considered by
Maloney (1996), the medium would be molecular only above a critical
pressure, depending on the radiation pressure, and could be atomic
at all radii.

In the case of four of our sources which have X-ray fluxes from the
Einstein survey listed by Wilkes et al.\ (1994) (PKS~0438$-$436,
0537$-$441, 1725+044, and 2223$-$052) the above condition may not be
satisfied. These sources have X-ray luminosities in the range
$5\times10^{44}-2\times10^{47}\ergs$ in the 0.2--4.5\,keV band (for an
assumed spectral slope $\alpha=1$; Wilkes et al.\ 1994) which largely
overlaps the hard X-ray band after allowing for the source redshifts.
Using similar values as Maloney et al.  for the other variables, these
X-ray luminosities imply that the ionization parameters may be high
enough to dissociate the CO. This is even more likely if we use the
lower values of the density ($10^6-10^8\cmcb$) used in the models of
Seyfert AGN tori by Krolik \& Lepp (1989).

\item 
There could also be CO present, but with a very low optical depth in
the lower transitions because of high excitation temperatures. The
upper limits in Table~2 were obtained using a temperature of 10\K, but
the column densities derived vary as $T_{ex}^{2}$, and could be 10$^4$
times higher if the excitation temperature of the molecular gas is
$\sim$ 1000K.
Close to the transition point between warm atomic and cool molecular gas
in a torus,
the kinetic temperature of of the molecular gas is $\sim$700\K (Maloney
1996). The high density, required for the existence of the molecular phase
(see above), ensures that the rotational transitions will be thermalized.
The excitation temperature will therefore be similar to the kinetic
temperature. The limits to the hydrogen column density derived from our
data, would then be higher than those derived from the optical reddening.

\item
Maloney et al.\ (1994) showed how the CO molecules could be radiatively
coupled to the strong radio continuum source, increasing the rotation
temperature and lowering the fractional population at low J--levels and,
hence, its optical depth.
This effect may cause the lack of molecular absorption in Cygnus~A with
its exceptionally luminous radio emission, but is not likely to be
generally important in radio galaxies (Maloney 1996).

\item 
The absorption may not take place at the AGN redshift, but in
intervening systems at lower redshifts. Whilst this possibility is
perhaps the most consistent with our non-detection of molecular
absorption, Webster et al.\ (1995) argue that intrinsic absorption is
more likely because of the lack of dependence with redshift.

\item 
We should mention the possibility that the reddening is not real, but
the result of continuum emission that is intrinsically red. Radio
galaxies are known to have red optical to infra-red colours, so the red
colours might be due to light from the quasar host galaxies. However at
the redshifts we are considering, typical radio galaxies are too faint
($K>18$ for $z>0.5$) to have any significant contribution to the
colours (see Lilly \& Longair 1984). In a recent comparison with X-ray
selected and optically-selected quasar samples, Boyle \& di Matteo
(1996) conclude that most of the spread in the colours is due to
effects other than dust obscuration.

\end{enumerate}

In summary, we found no evidence for molecular absorption at the
emission redshift of the 13 AGN in our sample. Several of the sources
have significant X-ray absorption or are optically reddened, implying a
high column density of absorbing gas. Our results imply that either the
absorption takes place along the line-of-sight to the AGN outside the
host galaxy, or that the absorbing material is hot or not associated
with molecular species like CO. In the model where the absorbing
material is concentrated in a parsec-scale dusty torus, we suggest that
the molecular gas is dissociated by the X-ray flux from the nucleus.
This in principle allows us to put upper limits on the density of the
torus, although we defer a detailed analysis until we can measure a
larger sample of sources.

\acknowledgements

We wish to thank Phil Maloney (the referee) and Ski Antonucci for
helpful suggestions about this manuscript. We also thank Belinda Wilkes
and Paolo Ciliegi for additional comments and Wolfgang Brinkmann for
making the ROSAT $N_H$ estimates for us. Rachel Webster kindly provided
the $Kn$ data in advance of publication and Richard Elston kindly made
an additional $Ks$ measurement for us.  We are very grateful to the
IRAM engineers and operators for the impressive work in tuning many
different frequencies during our runs as well as the efficient support
by the SEST team.  MJD would like to acknowledge travel funding from
the Australia-France Cooperation in Astronomy grant, as well as the
hospitality of the Observatoire de Paris during a visit.  This work has
made use of the NASA/IPAC Extragalactic Database (NED) which is
operated by the Jet Propulsion Laboratory, Caltech, under contract with
the National Aeronautics and Space Administration.

%_____________________________________________________________________


\begin{thebibliography}{}
%
\bibitem{} Antonucci, R. 1993, ARA\&A 31, 473
\bibitem{} Antonucci, R., Miller, J. 1985, ApJ 297, 621
\bibitem{} Arnaud, K.A., Johnstone, R.M., Fabian, A.C., Crawford, C.S., 
Nulsen, P.E.J., Shafer, R.A., Mushotzky, R.F. 1987, MNRAS, 227, 241
\bibitem{} Baldwin, J.A., Wampler, E.J., Burbidge, E.M. 1981, ApJ, 243, 76
\bibitem{} Barvainis, R., Antonucci, R. 1994, AJ, 107, 1291
\bibitem{} Bergman, P., Aalto, S., Black, J.H., Rydbeck, G. 1992, A\&A 265, 403
\bibitem{} Bohlin, R.C., Savage, B.D., Drake, J.F. 1978, ApJ, 224, 132
\bibitem{} Boyle, B.J., di Matteo, T. 1996, MNRAS, in press
\bibitem{} Carilli, C.L., Perley, R.A., Harris, D.E. 1994, MNRAS, 270, 173
\bibitem{} Combes, F., Wiklind, T.  1995,  A\&A 303, L61    
\bibitem{} Conway, J.E., Blanco, P.R. 1995, ApJ 449, L131
\bibitem{} Dickman, R.L. 1978, ApJS, 37, 407
\bibitem{} Drinkwater, M. J., Savage, A., Webster, R.L., Condon, J.J., 
Ellison, S.L., Francis, P.J., Jauncey, D.L., Lovell, J., Peterson, B.A. 
1996, MNRAS, in press
\bibitem{} Elvis, M., Fiore, F., Wilkes, B., McDowell, J., Bechtold, J. 
1994a, ApJ 422 60
\bibitem{} Elvis, M., Fiore, F., Mathur, S., Wilkes, B.J. 1994b, ApJ, 425, 103
\bibitem{} Israel, F.P., van Dishoeck, E.F., Baas, F., Koorneef, J., Black, 
J.H., de Graauw, T. 1990, A\&A 227, 342
\bibitem{} Krolik, J.H., Lepp, S. 1989, ApJ, 347, 179
\bibitem{} Lilly, S.J., Longair, M.S. 1984, MNRAS, 211, 833
\bibitem{} Maloney, P.R. 1996, in "Cygnus A -- Study of a Radio Galaxy", 
eds. C.L. Carilli and D.E. Harris (Cambridge: Cambridge
University Press), in press.
\bibitem{} Maloney, P.R., Begelman, M.C., Rees, M.J. 1994, ApJ 432, 606
\bibitem{} Miyoshi, M., Moran, J., Herrnstein, J., Greenhill, L., Nakai, N., 
Diamond, P., Inoue, M. 1995, Nature, 373, 127
\bibitem{} Morganti, R., Killeen, N.E.B., Tadhunter, C.N. 1993, MNRAS, 263, 1023
\bibitem{} Planesas, P., Scoville, N., Myers, S.T. 1991, ApJ 369, 364
\bibitem{} Rydbeck, G., Wiklind, T., Cameron, M., Wild, W., Eckart, A., 
Genzel, R., Rothermel, H. 1993, A\&A 270, L13
\bibitem{} Schartel, N., Fink, H., Brinkmann, W., Truemper, J. 1992, in  X 
Ray Emission from Active Galactic Nuclei and the Cosmic X Ray Background, 
eds. W. Brinkmann and J. Truemper (Garching: Max-Planck-Institut f\"ur
extraterrestrische Physik), p. 195 %  1992xeag.rept..195S
\bibitem{} Stark, A.A., Gammie, C.F., Wilson, R.W., Bally, J., Linke, R.A., 
Heiles, C., Hurwitz, M. 1992, ApJS, 79, 77
\bibitem{} Webster, R.L., Francis, P.J., Peterson, B.A., Drinkwater, M.J., 
Masci, F.J. 1995, Nature, 375, 469
\bibitem{} Wiklind, T., Combes, F. 1994, A\&A 286, L9    % PKS~1413+135
\bibitem{} Wiklind, T., Combes, F. 1995, A\&A 299, 382   % B0218+357
\bibitem{} Wiklind, T., Combes, F. 1996a, Nature, 379, 139 % PKS1830-211
\bibitem{} Wiklind, T., Combes, F. 1996b, A\&A submitted  % AGN 1504+377
\bibitem{} Wilkes, B.J., Tananbaum, H., Worrall, D.M., Avni, Y., Oey, M.S., 
Flanagan, J. 1994, ApJS, 92, 53
\end{thebibliography}
\end{document}